\documentclass[aps,prb,twocolumn,superscriptaddress,showpacs,floatfix]{revtex4}

\usepackage{graphicx}
\usepackage{dcolumn}
\usepackage{bm}
\usepackage{stmaryrd}
\usepackage{latexsym}
\usepackage{amssymb}
\usepackage{amsfonts}
\usepackage{amsmath}
\usepackage{color}

\begin{document}
\title{Spin-drag relaxation time in one-dimensional spin-polarized Fermi gases}
\date{\today}

\author{Diego Rainis}
\affiliation{NEST-CNR-INFM and Scuola Normale Superiore, I-56126 Pisa, Italy}
\author{Marco Polini}
\email{m.polini@sns.it}
\affiliation{NEST-CNR-INFM and Scuola Normale Superiore, I-56126 Pisa, Italy}
\author{M.P. Tosi}
\affiliation{NEST-CNR-INFM and Scuola Normale Superiore, I-56126 Pisa, Italy}
\author{G. Vignale}
\affiliation{Department of Physics and Astronomy, University of
Missouri, Columbia, Missouri 65211, USA}

\begin{abstract}
Spin propagation in systems of one-dimensional interacting fermions at finite temperature is intrinsically diffusive. The spreading rate of a spin packet is controlled by a transport coefficient termed ``spin drag" relaxation time $\tau_{\rm sd}$. In this paper we present both numerical and analytical calculations of $\tau_{\rm sd}$ for 
a two-component spin-polarized cold Fermi gas trapped inside a tight atomic waveguide. At low temperatures we find an activation law for $\tau_{\rm sd}$, in agreement with earlier calculations of Coulomb drag between slightly asymmetric quantum wires, but with a different and much stronger temperature dependence of the prefactor. Our results provide a fundamental input for microscopic time-dependent spin-density functional theory calculations of spin transport in $1D$ inhomogeneous systems of interacting fermions.
\end{abstract}
\maketitle

\maketitle

\section{Introduction}
\label{sect:intro}

Quantum many-body systems of one-dimensional ($1D$) interacting particles have attracted an enormous interest for more than fifty years~\cite{giamarchi_book} and are nowadays available in a large number of different laboratory systems ranging from single-wall carbon nanotubes~\cite{Saito_book} to semiconductor nanowires~\cite{Tans_1998}, conducting molecules~\cite{Nitzan_2003}, chiral Luttinger liquids at fractional quantum Hall edges~\cite{xLL}, and trapped atomic gases~\cite{moritz_2005}.  

Regardless of statistics, the effective low-energy description of these systems is based on a harmonic theory of long-wavelength fluctuations~\cite{haldane}, {\it i.e.} on the ``Luttinger liquid" model~\cite{giamarchi_book}. The distinctive feature of the Luttinger liquid is that its low-energy excitations are independent collective oscillations of the charge density {\it or} the spin density, as opposed to individual quasiparticles  that carry both charge and spin.  
This leads immediately to the phenomenon of spin-charge separation~\cite{giamarchi_book}, {\it i.e.} the fact that the low-energy spin and charge excitations of $1D$ interacting fermions are completely decoupled and propagate with different velocities. This behavior has been recently demonstrated by Kollath {\it et al.}~\cite{kollath_prl_2005} in a numerical time-dependent density-matrix renormalization-group study of the $1D$ Hubbard model for fermions, by Kleine {\it et al.}~\cite{kleine_2007}  in a similar study of the two-component Bose-Hubbard model, and, analytically, by Kecke {\it et al.}~\cite{kecke_prl_2005} for interacting fermions in a  $1D$ harmonic trap.  The possibility of studying these phenomena experimentally in $1D$ two-component cold Fermi gases~\cite{moritz_2005},  where ``spin" and ``charge" refer, respectively,  to two internal (hyperfine) atomic states and to the atomic mass density, was first highlighted by Recati {\it et al.}~\cite{recati_prl_2003}. 


In a recent paper~\cite{polini_PRL_2007} two of us have pointed out a new aspect of spin-charge separation: namely, spin excitations are intrinsically damped at finite temperature, while charge excitations are not. 
The physical reason for  this difference is easy to grasp. 
In a pure spin pulse the up-spin and down-spin components of the current are always equal and oppositely directed, 
so that the density remains constant. 
The relative motion of the two components gives rise to a form of friction known, in electronic systems,  as ``{\it spin Coulomb drag}"~\cite{scd_giovanni,scd_flensberg,scd_tosi,weber_nature_2005}. 
This is responsible for a hydrodynamic behavior due to a randomization of ``spin momentum" deriving from excitations in both the $q\to 0$ and $q\to 2 k_{\rm F}$ regions. In the course of time this leads to diffusive spreading of the packet. No such effect is present in the propagation of  pure density pulses, which are therefore essentially free of diffusion in the long wavelength limit~\cite{pustilnik_2006} (by contrast, a density pulse in a normal Fermi liquid is always expected to decay into particle-hole pairs -- a process known as Landau damping~\cite{Giuliani_and_Vignale}). 

The analysis of Ref.~\onlinecite{polini_PRL_2007} was limited to spin-compensated (unpolarized) systems.  In particular, the calculation of the spin-drag relaxation time $\tau_{\rm sd}$ was done only for unpolarized systems. 
It is of great interest to extend the calculation to spin-polarized systems.
First, a finite degree of spin-polarization is the most general situation in experiments. 
Second, and most important, any attempt to study the dynamics of spin packets beyond the linear approximation will have to take into account the spin-polarization of the liquid within the packet. This is particularly crucial in the time-dependent spin-current-density functional approach~\cite{Qian03}, which treats spin-current relaxation through a local spin-drag relaxation rate. Calculations done by this method will therefore require knowledge of $\tau_{\rm sd}$ as a function of the local spin polarization and temperature. 
Finally, our study is relevant to the closely related problem of the regular Coulomb drag (no spin involved) between two semiconductor wires with different carrier densities. Here the two wires play formally the role of the two spin orientations, and the difference in density corresponds to the spin polarization.  Our careful analysis of the low temperature regime yields a new formula for the temperature and polarization dependence of $\tau_{\rm sd}$, which is completely different from the one that was obtained previously by a qualitative method~\cite{pustilnik_2003}.

The contents of the paper are briefly described as follows. In Sect.~\ref{sect:theory} we introduce the model Hamiltonian and we present our numerical calculations of the spin-drag relaxation rate of $1D$ spin-polarized Fermi gases. In Sect.~\ref{sect:analytical} we report and discuss our main analytical results. Finally, in Sect.~\ref{sect:conclusion} we summarize our main conclusions.

\section{Spin-drag relaxation time for spin-polarized systems}
\label{sect:theory}

We consider a two-component Fermi gas with $N$ atoms confined inside a tight atomic waveguide of length $L$ (an ``atomic quantum wire") along the $x$ direction, realized {\it e.g.} using two overlapping standing waves along the $y$ and the $z$ axis as in Ref.~\onlinecite{moritz_2005}. The atomic waveguide provides a tight harmonic confinement in the $y-z$ plane~\cite{footnote_1} characterized by a large trapping frequency $\omega_\perp\simeq 2\pi\times 10~{\rm kHz}$. The two species of fermionic atoms are assumed to have the same mass $m$ and different spin  $\sigma$, $\sigma=\uparrow$ or $\downarrow$. The number of fermions with spin $\uparrow$ ($N_\uparrow$) is taken to be larger than the number of fermions with spin $\downarrow$ ($N_\downarrow$), $N_\uparrow > N_\downarrow$. The fermions have quadratic dispersion,  $\varepsilon_k=\hbar^2k^2/(2m)$,  and interact  {\it via} a zero-range $s$-wave potential $v(x)=g_{\rm 1D}\delta(x)$. The Fourier transform of such a real-space potential is a simple constant, $v_q=g_{\rm 1D}$. The system is thereby governed by the Yang Hamiltonian~\cite{yang_1967}
\begin{eqnarray}\label{eq:gy_momentum}
{\cal H}=-\frac{\hbar^2}{2m}\sum_{i=1}^{N}\frac{\partial^2}{\partial x^2_i}+g_{\rm  1D}\sum_{i=1}^{N_{\uparrow}}\sum_{j=1}^{N_{\downarrow}}\delta(x_i-x_j)\,.
\end{eqnarray}
The effective $1D$ coupling constant $g_{\rm 1D}$ can be tuned~\cite{moritz_2005} by using a magnetic field-induced Feshbach resonance to change the $3D$ scattering length $a_{\rm 3D}$. In the limit $a_{\rm 3D}\ll a_\perp$, where $a_\perp=\sqrt{\hbar^2/(m\omega_\perp)}$, one finds $g_{\rm 1D}=2\hbar^2 a_{\rm 3D}/(ma^2_\perp)$~\cite{olshanii_1998}. In the thermodynamic limit ($N_\sigma,L\rightarrow \infty, N_\sigma/L=n_\sigma$) the properties of the system described by ${\cal H}$ are determined by the linear density $n=n_\uparrow+n_\downarrow$, by the degree of spin polarization $\zeta=(N_\uparrow-N_\downarrow)/N$, and by the effective coupling $g_{\rm 1D}$. For future purposes it will be useful to introduce the dimensionless Yang interaction parameter $\gamma=m g_{\rm 1D}/(\hbar^2 n)$. We also introduce the Fermi wave number $k_{\rm F}=\pi n/2$ of the unpolarized system, the Fermi velocity $v_{\rm F}=\hbar k_{\rm F}/m$, and the Fermi energy $\varepsilon_{\rm F}=\hbar^2 k^2_{\rm F}/(2m)$.
\begin{figure}
\begin{center}
\includegraphics[width=1.0\linewidth]{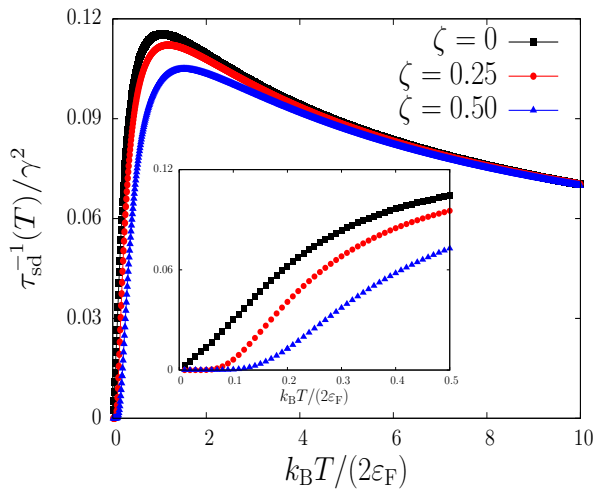}
\caption{(Color online) Spin-drag relaxation rate $\tau^{-1}_{\rm sd}$ (in units of $\varepsilon_{\rm  F}/\hbar$ and divided by $\gamma^2$) as a function of the reduced temperature $k_{\rm B} T/(2\varepsilon_{\rm F})$ for three different values of the spin polarization $\zeta$. In the inset we show a zoom of the low-temperature region $0\leq k_{\rm B}T/(2\varepsilon_{\rm F}) \leq 0.5$.}
\label{fig:one}
\end{center}
\end{figure}

Within second-order perturbation theory the spin-drag relaxation rate (at zero frequency) is given by the formula~\cite{scd_giovanni}
\begin{eqnarray}\label{eq:nd_order_tau}
\frac{1}{\tau_{\rm sd}}&=&\frac{\hbar^2 n}{n_{\uparrow}n_{\downarrow} m k_{\rm B}T}\int_{0}^{+\infty}\frac{dq}{2\pi}~q^2v_q^2\nonumber\\
&\times&\int_{0}^{+\infty}\frac{d\omega}{\pi}~\frac{\Im m \chi^{(0)}_{\uparrow}(q,\omega)\Im m\chi^{(0)}_{\downarrow}(q,\omega)}
{\sinh^2[\hbar\omega/(2k_{\rm B}T)]}\,,
\end{eqnarray}
where $\Im m\chi^{(0)}_{\sigma}(q,\omega)$ is the spin-resolved finite-temperature expression for the imaginary part of the $1D$ Lindhard function~\cite{Giuliani_and_Vignale},
\begin{widetext}
\begin{equation}\label{eq:imchi0}
\Im m\chi^{(0)}_{\sigma}(q,\omega) = -\frac{\pi k_{{\rm F}\sigma}}{2q} N_{\sigma}(0) \left\{
\frac{1}{{\rm exp}\{[\nu^2_{-, \sigma}\varepsilon_{{\rm F}\sigma} - \mu_{\sigma}(T)]/(k_{\rm B} T)\} + 1} -
\frac{1}{{\rm exp}\{[\nu^2_{+, \sigma}\varepsilon_{{\rm F}\sigma} - \mu_{\sigma}(T)]/(k_{\rm B} T)\} + 1} 
\right\}\,.
\end{equation}
\end{widetext}
Here 
$k_{{\rm F}\sigma} = k_{\rm F} [1+{\rm sgn}(\sigma)\zeta]$ [by definition 
${\rm sgn}(\sigma)=+1$ for $\sigma=\uparrow$ and to $-1$ for $\sigma=\downarrow$], 
$N_\sigma(0) = m/(\pi \hbar^2 k_{{\rm F}\sigma})$ is the spin-resolved density of states at the Fermi level, $\varepsilon_{{\rm F}\sigma}=\varepsilon_{\rm F}[1+{\rm sgn}(\sigma)\zeta]^2$, and
\begin{equation}\label{eq:nus}
\nu_{\pm, \sigma} = \frac{m\omega}{q\hbar k_{{\rm F}\sigma}}\pm \frac{q}{2k_{{\rm F}\sigma}}\,.
\end{equation}
In Eq.~(\ref{eq:imchi0}) $\mu_\sigma(T)$ is the spin-resolved chemical potential, which is determined by the normalization condition
\begin{equation}\label{eq:chem_pot}
n_\sigma =\int_{-\infty}^{\infty}\frac{dk}{2\pi}
\frac{1}{\exp\{[\varepsilon_k - \mu_\sigma(T)]/(k_{\rm B} T)\} + 1}\,.
\end{equation}

In Fig.~\ref{fig:one} we report some illustrative numerical results for the spin-drag relaxation rate 
$\tau^{-1}_{\rm sd}$ of a $1D$ spin-polarized Fermi gas as calculated from Eqs.~(\ref{eq:nd_order_tau})-(\ref{eq:chem_pot}). 
Note from the inset in Fig.~\ref{fig:one} 
that a finite value of $\zeta$ has a dramatic effect on the low-temperature behavior of the spin-drag relaxation rate, changing it from linear~\cite{polini_PRL_2007} to exponentially activated. At high temperatures $\tau^{-1}_{\rm sd}$ is seen to be insensitive to the degree of spin polarization, becoming asymptotically equal to the unpolarized $\zeta=0$ result. Both observations will be demonstrated in the next Section with analytical calculations. 

Note finally that $\tau^{-1}_{\rm sd}$ is largest for the unpolarized case, simply because the ``overlap" between $\Im m\chi^{(0)}_{\uparrow}(q,\omega)$ and $\Im m\chi^{(0)}_{\downarrow}(q,\omega)$ is maximum at $\zeta=0$.

\section{Analytical results}
\label{sect:analytical}

Low- and high-temperature analytical expressions for $\tau^{-1}_{\rm sd}$ have been derived in the unpolarized case in Ref.~\onlinecite{polini_PRL_2007}. 
Here we proceed to derive some analytical results for $\tau^{-1}_{\rm sd}$ for $\zeta\neq 0$. Similar calculations have been performed earlier by Pustilnik {\it et al.}~\cite{pustilnik_2003} with the aim of studying 
the Coulomb drag between slightly asymmetric quantum wires. We will comment at length below 
on the connection between our calculations and the calculations reported in Ref.~\onlinecite{pustilnik_2003}.

It is crucial to realize~\cite{pustilnik_2003} that a finite degree of spin polarization implies the existence of a new temperature scale ({\it i.e.} in addition to the Fermi temperature $T_{\rm F}$), that we define as $T_\zeta \equiv T_{{\rm F}\uparrow}-T_{{\rm F}\downarrow}=4\zeta T_{\rm F}$, where $T_{{\rm F}\sigma}=\varepsilon_{{\rm F}\sigma}/k_{\rm B}$. In what follows we will derive analytical results for $T \ll {\rm min}(T_\zeta, T_{{\rm F}\downarrow})$. Note that this inequality guarantees also that the minority $\downarrow$-spin component is always in the regime of quantum degeneracy, thus allowing us to use 
Eq.~(\ref{eq:imchi0}) also for the imaginary part of the spin-resolved density-density response function for the minority spin population.

Before proceeding to illustrate our analytical procedure [outlined from Eq.~(\ref{eq:sd_2}) down below], it is instructive to summarize the main steps followed by Pustilnik {\it et al.} to derive Eq.~(16) in Ref.~\onlinecite{pustilnik_2003}. As it will be clear below, there are two contributions to the inverse spin-drag relaxation time, one coming from the low-$q$ region and the other one coming from the $q\approx 2k_{\rm F}$ region, similarly to what happens in the unpolarized case~\cite{polini_PRL_2007}. In Ref.~\onlinecite{pustilnik_2003} the authors focus only on the low-$q$ contribution and use the $T=0$ rectangular form for the imaginary part of the Lindhard response function, {\it i.e.} $\Im m \chi_\sigma^{(0)}(q,\omega)|_{T=0}=[m/(2\hbar^2 q)]\Theta(\hbar q^2/(2m)-|\omega-v_{{\rm F}\sigma}q|)$, where $\Theta(x)$ is the Heaviside step function. This immediately implies that the product $\Im m \chi^{(0)}_{\uparrow}(q,\omega)\Im m\chi^{(0)}_{\downarrow}(q,\omega)$ in Eq.~(\ref{eq:nd_order_tau}) is itself a rectangle of width $\delta \omega=\hbar q^2/m-2\zeta \hbar q k_{\rm F}/m$ centred at $\omega=qv_{\rm F}$.  For this reason Pustilnik {\it et al.}~\cite{pustilnik_2003} 
substitute the thermal factor $1/\sinh^2[\hbar\omega/(2k_{\rm B}T)]$ with $1/\sinh^2[\hbar q v_{\rm F}/(2k_{\rm B}T)]$, bringing it outside the $\omega$ integration in Eq.~(\ref{eq:nd_order_tau}). Then, using the fact that
\begin{eqnarray}
\alpha(q,\zeta)&=&\left.\int_0^{+\infty}d\omega~\Im m \chi^{(0)}_{\uparrow}(q,\omega)\Im m\chi^{(0)}_{\downarrow}(q,\omega)\right|_{T=0}\nonumber\\
&=&\frac{m}{4\hbar^3q}(q-2\zeta k_{\rm F})\Theta(q-2\zeta k_{\rm F})\,,
\end{eqnarray}
they arrive  at the result~\cite{pustilnik_2003} 
\begin{eqnarray}\label{eq:pustilnik}
\left.\frac{1}{\tau_{\rm sd}}\right|_{\rm PMGA}\!\!\!& = & \!\frac{1}{2\hbar \pi^2 n k_{\rm B}T}
\int_{2\zeta k_{\rm F}}^{\infty}dq~v^2_q 
\frac{q(q-2\zeta k_{\rm F})}{\sinh^2[\hbar q v_{\rm F}/(2k_{\rm B}T)]}\nonumber\\
& \propto & \zeta T~\exp{(-T_\zeta/T)}\,,
\end{eqnarray}
for $\zeta\ll 1$.

\begin{figure}
\begin{center}
\includegraphics[width=0.75\linewidth]{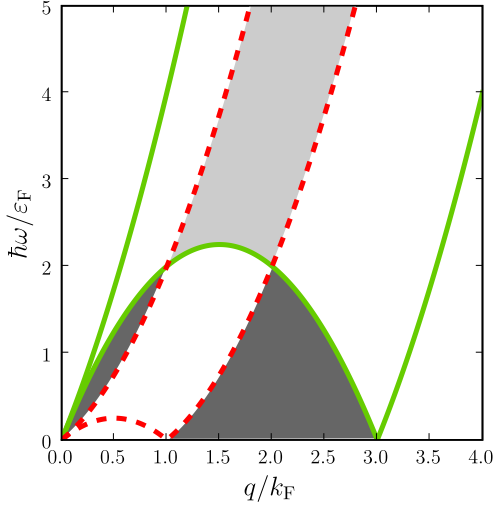}
\caption{(Color online) The one-dimensional particle-hole continuum of the majority spin-$\uparrow$ component (solid lines) and the particle-hole continuum of the minority spin-$\downarrow$ component (dashed lines). The dark grey regions are the integration regions that we have considered in our calculations. The light grey region is the one considered by Pustilnik {\it et al.}~\cite{pustilnik_2003}. The small thermal-overlap-induced value of $\Im m \chi^{(0)}_{\uparrow}(q,\omega)\times \Im m \chi^{(0)}_{\downarrow}(q,\omega)$ in the dark regions is more than compensated by the large value of $1/\sinh^2[\hbar\omega/(2k_{\rm B}T)]$.}
\label{fig:two}
\end{center}
\end{figure}

Using the zero-temperature expression for the product $\Im m \chi^{(0)}_{\uparrow}(q,\omega)\Im m\chi^{(0)}_{\downarrow}(q,\omega)$ results, however, in the wrong pre-exponential factor: the thermal tails of $\Im m \chi_\sigma^{(0)}(q,\omega)$ are exponentially small but they have a finite overlap for the two spin populations precisely in the low-frequency region where the factor $1/\sinh^2[\hbar\omega/(2k_{\rm B}T)]$ is large and thus can originate contributions
that are much larger, in the low-temperature limit, than the contributions retained in Eq.~(\ref{eq:pustilnik}).  The two regions where the thermal tails of 
$\Im m \chi_{\uparrow}^{(0)}(q,\omega)$ and $\Im m \chi_{\downarrow}^{(0)}(q,\omega)$ overlap are depicted as dark grey regions in Fig.~\ref{fig:two} and will be discussed at length below. In what follows we will show that the functional dependence on $\zeta$ and $T$ of the pre-exponential factor in Eq.~(\ref{eq:pustilnik}) is incorrect. In particular the prefactor $T$ should be replaced by the much larger $T^{-1}$. To show this, we proceed as follows.

\begin{figure}
\begin{center}
\includegraphics[width=1.0\linewidth]{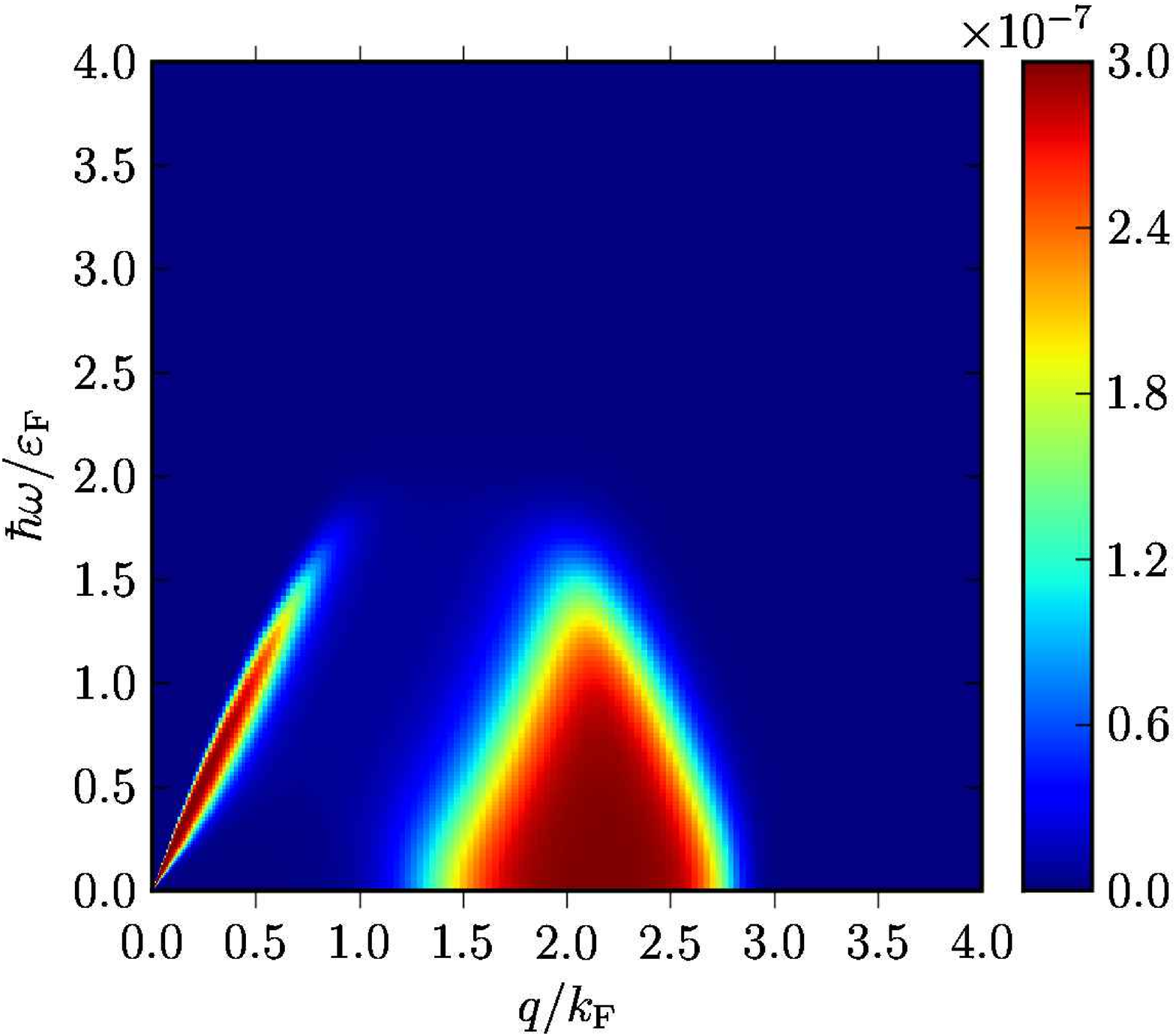}
\includegraphics[width=1.0\linewidth]{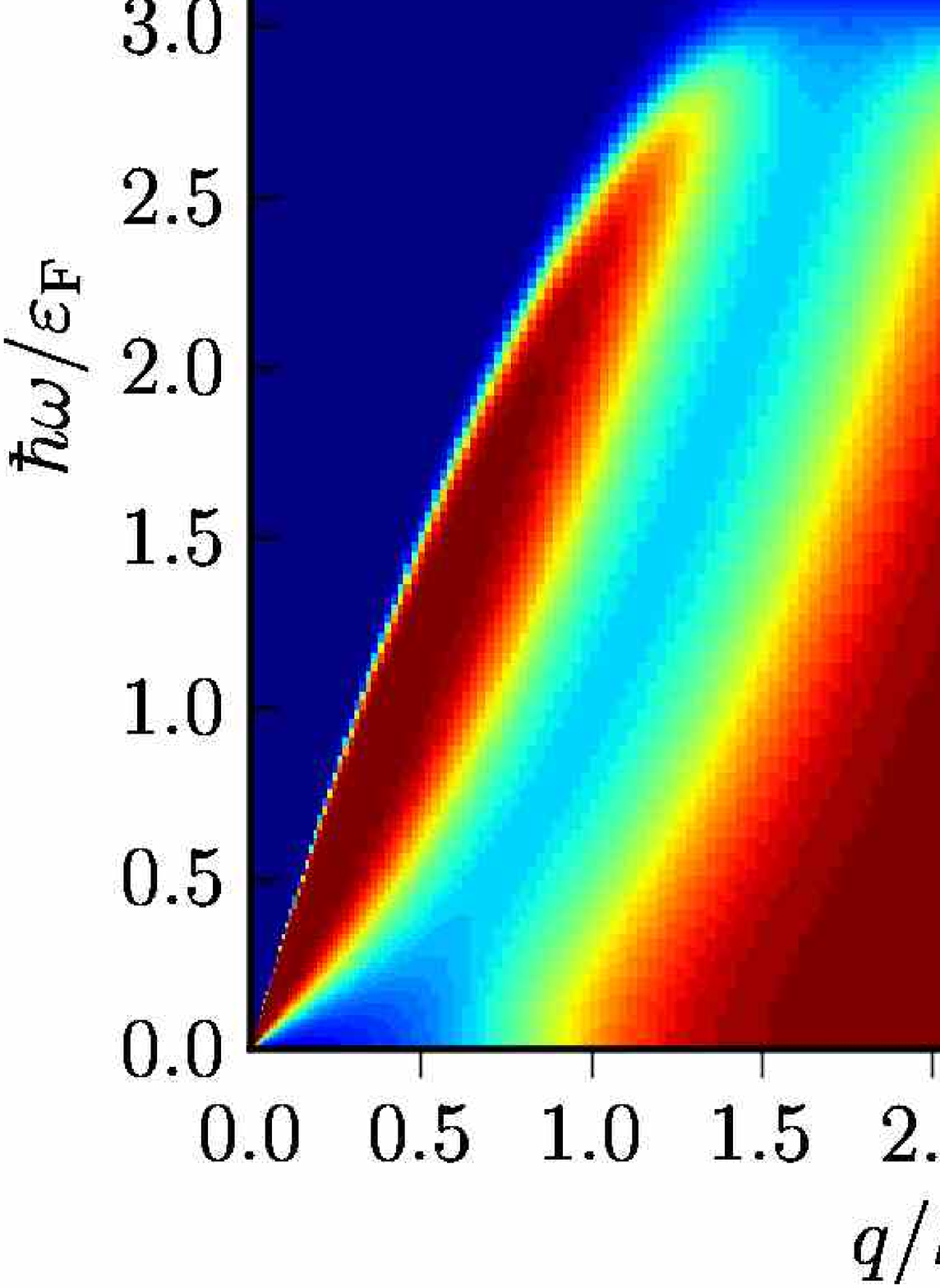}
\caption{(Color online) A color density plot of $I'(q,\omega)$ as a function of $q/k_{\rm F}$ and $\hbar\omega/\varepsilon_{\rm F}$ for $T=0.1~T_{\rm F}$. Top panel: $\zeta=0.4$. Bottom panel: $\zeta=0.75$. The two regions, olive leaf and shark fin, are clearly distinguishable.}
\label{fig:three}
\end{center}
\end{figure}

\begin{figure}
\begin{center}
\includegraphics[width=1.0\linewidth]{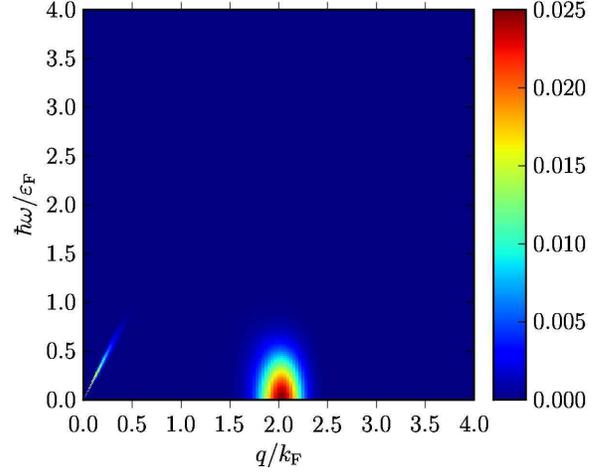}
\caption{(Color online) Same as in Fig.~\ref{fig:three} but for $\zeta=0.1$.}
\label{fig:four}
\end{center}
\end{figure}

Inserting Eq.~(\ref{eq:imchi0}) in Eq.~(\ref{eq:nd_order_tau}) we find, after some straigthforward algebraic manipulations, 
the following expression for $\tau^{-1}_{\rm sd}$,
\begin{equation}\label{eq:sd_2}
\frac{1}{\tau_{\rm sd}}=\frac{mn \beta}{8\pi^2\hbar^2n_\uparrow n_\downarrow} 
\int_{0}^{\infty}dq~v^2_q\int_{0}^{+\infty}d\omega~\frac{I(q,\omega)}{\sinh^2(\beta\hbar\omega/2)}\,.
\end{equation}
Here $\beta=(k_{\rm B} T)^{-1}$ and $I(q,\omega)$ can be written as
\begin{widetext}
\begin{eqnarray}\label{eq:I}
I(q,\omega)=\frac{\sinh^2(\beta\hbar\omega/2)}{\cosh(\beta \theta_1)/2 + \cosh(\beta \theta_2)/2 + 
\cosh^2(\beta\hbar\omega/2) + 
2 \cosh(\beta \theta_1/2)\cosh(\beta \theta_2/2)\cosh(\beta\hbar\omega/2)}
\end{eqnarray}
\end{widetext}
with
\begin{eqnarray}
\left\{
\begin{array}{l}
{\displaystyle \theta_1= 4\zeta \varepsilon_{\rm F}}\vspace{0.1 cm}\\
{\displaystyle \theta_2=\frac{m\omega^2}{q^2} + \frac{\hbar^2 q^2}{4m} -2\varepsilon_{\rm F}(1+\zeta^2)}
\end{array}
\right.\,.
\end{eqnarray}
We would like to stress at this point that Eq.~(\ref{eq:sd_2}) has been obtained from Eq.~(\ref{eq:nd_order_tau}) without any approximation. 
Two facts are remarkable.  To begin with, we note that, quite surprisingly, an {\it exact} cancellation occurs between the factor $\sinh^2(\beta\hbar\omega/2)$ in  the numerator of Eq.~(\ref{eq:I}) and the same factor in the denominator of the $\omega$-integrand in Eq.~(\ref{eq:sd_2}). Secondly, a new temperature scale, $T_\zeta=4\zeta \varepsilon_{\rm F}/k_{\rm B}$, has set in. A color plot of the function $I(q,\omega)/\sinh^2(\beta\hbar\omega/2)$, {\it i.e.} the integrand in Eq.~(\ref{eq:sd_2}), is reported in Fig.~\ref{fig:three} for two values of $\zeta$ at $T=0.1~T_{\rm F}$. We can thus distinguish clearly two regions in the $(q,\omega)$ plane where this function is non-zero: one region, ``close" to $q=0$, which has the shape of an olive leaf and another region, ``close" to $2k_{\rm F}$, which has the shape of a shark fin. These two regions will be defined accurately in what follows. Note also that on decreasing $\zeta$ the area of each of these regions decreases (see also Fig.~\ref{fig:four}): 
in the limit $\zeta\to 0$ the two regions become two points located at $(0,0)$ and $(2k_{\rm F},0)$ in the $(q,\omega)$ plane, as expected from the calculations performed in Ref.~\onlinecite{polini_PRL_2007}.

Thanks to the aforementioned cancellation, to make some analytical progress in the calculation of $\tau_{\rm sd}$ for $T\to 0$ we just need to understand what happens for $T\to 0$ to the four $\cosh$ terms in the denominator of Eq.~(\ref{eq:I}). To begin with, note that the first term 
$\cosh(\beta\theta_1)/2$ is independent of $q$ and $\omega$. In the limit $T \to 0$ we can substitute all hyperbolic cosine functions with simple exponentials,
\begin{widetext}
\begin{eqnarray}\label{eq:Ibis}
I'(q,\omega)&\equiv&\frac{I(q,\omega)}{\sinh^2(\beta\hbar\omega/2)}\stackrel{T\to 0}{\to} \frac{4}{\exp(\beta \theta_1) + \exp(\beta |\theta_2|) + 
\exp(\beta\hbar\omega) + \exp[\beta (\theta_1+|\theta_2|+\hbar\omega)/2]}\nonumber\\
&=&\frac{4\exp(-\beta \theta_1)}{1+ \exp[\beta (|\theta_2|-\theta_1)] + 
\exp[\beta(\hbar\omega-\theta_1)] + \exp[\beta (|\theta_2|+\hbar\omega-\theta_1)/2]}\,,
\end{eqnarray}
\end{widetext}
where we have taken into account the possibility that $\theta_2$ becomes negative [contrary to the other arguments of the $\cosh$ functions in Eq.~(\ref{eq:I}) that are positive definite].  In the limit $T\to 0$ the exponentials in the denominator of the second line of Eq.~(\ref{eq:Ibis}) are either zero (if their argument is negative) or infinity (if their argument is positive).  This implies that  $\exp(\beta \theta_1)I'(q,\omega)$ differs from zero only when the arguments of all the exponentials in the denominator of Eq.~(\ref{eq:Ibis}) are negative.  Thus $I'(q,\omega)$  reduces to a very simple form,
\begin{eqnarray}
I'(q,\omega)&\stackrel{T\to 0}{\simeq}&4\exp(-\beta \theta_1)\Theta(\theta_1-|\theta_2|)\Theta(\theta_1-\hbar\omega)\nonumber\\
&\times&\Theta(\theta_1-|\theta_2|-\hbar\omega)\,.
\end{eqnarray}
More explicitly, $I'(q,\omega)\stackrel{T\to 0}{\simeq}4\exp(-\beta \theta_1)$ if $\theta_1-|\theta_2|-\hbar\omega>0$, {\it i.e.}
\begin{equation}\label{eq:regions}
\left|\frac{m\omega^2}{q^2} + \frac{\hbar^2 q^2}{4m} -2\varepsilon_{\rm F}(1+\zeta^2)\right|+\hbar\omega<4\zeta \varepsilon_{\rm F}\,,
\end{equation}
and zero elsewhere. Eq.~(\ref{eq:regions}) determines two disconnected regions in the $(q,\omega)$ plane where $I'(q,\omega)$ is non-zero: (i) a region that we have called above ``olive leaf" (${\cal D}_{\rm OL}$) enclosed between the two parabolic arches
\begin{equation}
\left\{
\begin{array}{l}
\omega_{\rm up}(q)=-\varepsilon_q/\hbar+\hbar k_{\rm F}(1+\zeta)q/m\\
\omega_{\rm down}(q)=\varepsilon_q/\hbar+\hbar k_{\rm F}(1-\zeta)q/m
\end{array}
\right.\,,
\end{equation}
with $0\leq q \leq 2\zeta k_{\rm F}$; and (ii) a region that we have called above ``shark fin" (${\cal D}_{\rm SF}$) enclosed between the two parabolic arches
\begin{equation}
\left\{
\begin{array}{l}
\omega_{\rm left}(q)=\varepsilon_q/\hbar-\hbar k_{\rm F}(1-\zeta)q/m\\
\omega_{\rm right}(q)=-\varepsilon_q/\hbar+\hbar k_{\rm F}(1+\zeta)q/m
\end{array}
\right.\,,
\end{equation}
with $2(1-\zeta)k_{\rm F} \leq q \leq 2(1+\zeta) k_{\rm F}$.  These regions have been shown in Fig.~\ref{fig:two}, 
{\it vis-\`a-vis} the region implicitly considered in Ref.~\onlinecite{pustilnik_2003}. Note that in these domains $I'(q,\omega)$ is a constant $=4\exp(-T_\zeta/T)$, independent of $q$ and $\omega$.

\begin{figure}
\begin{center}
\includegraphics[width=1.0\linewidth]{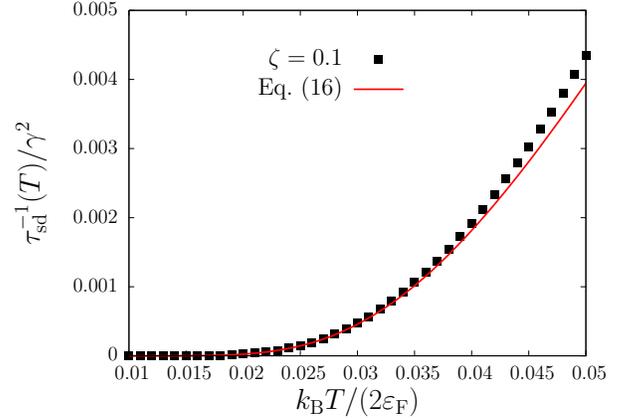}
\caption{(Color online) A comparison between the analytical formula in Eq.~(\ref{eq:sd_final}) (solid line) and the numerical results 
calculated from Eqs.~(\ref{eq:nd_order_tau})-(\ref{eq:chem_pot}) (filled squares).}
\label{fig:five}
\end{center}
\end{figure}

At this point we can easily calculate analytically the asymptotic $T\to 0$ behavior of the spin-drag relaxation rate 
in Eq.~(\ref{eq:sd_2}): for $v_q=g_{\rm 1D}$ we find
\begin{eqnarray}\label{eq:sd_final}
\frac{1}{\tau_{\rm sd}}&\stackrel {T\to 0}{\rightarrow}&\frac{4m}{8\pi^2\hbar^2n(1-\zeta^2)k_{\rm B}T} g^2_{\rm 1D}\times 4\exp(-T_\zeta/T)\nonumber\\
&\times&\int\int_{{\cal D}_{\rm OL}\cup{\cal D}_{\rm SF}}dqd\omega~\nonumber\\
&=&\left[\frac{32}{3\pi^3}\gamma^2~\frac{3\zeta^2+\zeta^3}{1-\zeta^2}~\frac{2\varepsilon_{\rm F}}{k_{\rm B}T}~
\exp{\left(-\frac{T_\zeta}{T}\right)}\right] \frac{\varepsilon_{\rm F}}{\hbar}\,,\nonumber\\
\end{eqnarray}
where we have used that $\int\int_{{\cal D}_{\rm OL}\cup{\cal D}_{\rm SF}}dqd\omega=(8\zeta^2+8\zeta^3/3)k_{\rm F}\varepsilon_{\rm F}/\hbar$ (sum of the two areas of the olive leaf and shark fin domains). In this equation the first contribution $\propto \zeta^2$ comes from the shark fin region ({\it i.e.} the region ``close" to $q=2k_{\rm F}$), while the second contribution $\propto \zeta^3$ comes from 
the olive leaf region ({\it i.e.} the region ``close" to $q=0$). The activation law $\exp{(-T_\zeta/T)}$  in Eq.~(\ref{eq:sd_final}) is identical to the one reported above in Eq.~(\ref{eq:pustilnik}) (see Ref.~\onlinecite{pustilnik_2003}). The prefactor, however, has a completely different functional dependence 
on temperature (proportional to $T^{-1}$ rather than $T$) and polarization.  
Quite interestingly, the scaling $\zeta \to T$ in Eq.~(\ref{eq:sd_final}) gives the correct temperature dependence of the unpolarized result, {\it i.e.} a linear term from $q \approx 2k_{\rm F}$ and a quadratic term from $q \approx 0$. In Fig.~\ref{fig:five} we show a comparison between the analytical formula (\ref{eq:sd_final}) and the numerical results for $\zeta=0.1$ as calculated from Eqs.~(\ref{eq:nd_order_tau})-(\ref{eq:chem_pot}). The agreement is clearly excellent in the asymptotic regime $T \ll T_\zeta$.

Before  concluding, we would like to calculate the high-temperature limit of the spin-drag relaxation rate. 
In this limit the spin-resoved chemical potential is simply given by $\mu_\sigma(T\to \infty)=k_{\rm B}T \ln[n_\sigma\lambda_{\rm dB}(T)]$, where $\lambda_{\rm dB}(T)=\sqrt{2\pi\hbar^2/(mk_{\rm B}T)}$ is the thermal de Broglie wavelength. Using this result one easily finds
\begin{eqnarray}\label{eq:high_Tb}
\frac{1}{\tau_{\rm sd}}&\stackrel{T\to \infty}{\to}&\frac{m}{2\hbar^3 n^2}
\frac{8}{\pi^{7/2}}\left(\frac{k_{\rm B} T}{2 \varepsilon_{\rm F}}\right)^{-3/2}\nonumber\\
&\times&\int_0^{+\infty}dq~qv^2_q \exp{\left(-\frac{\varepsilon_q}{2 k_{\rm B}T}\right)}\,,
\end{eqnarray}
which for $v_q=g_{\rm 1D}$ becomes
\begin{equation}\label{eq:unpolarized}
\frac{1}{\tau_{\rm sd}}\stackrel{T\to\infty}{\to}
 \frac{16}{\pi^{7/2}}\gamma^2\left(\frac{k_{\rm B}T}{2\varepsilon_{\rm F}}\right)^{-1/2}\frac{\varepsilon_{\rm F}}{\hbar}\,.
\end{equation}
At high-temperature the spin-drag relaxation rate becomes $\zeta$-independent and identical to the one calculated in Ref.~\onlinecite{polini_PRL_2007} for a strictly unpolarized system.

\section{Conclusions}
\label{sect:conclusion}

In summary, we have carefully studied within second-order perturbation theory the spin-drag relaxation time of a spin-polarized $1D$ Fermi gas. We obtain accurate numerical results for $1/\tau_{\rm sd}$ as a function of spin polarization and temperature (see Fig.~\ref{fig:one}), and also an accurate analytical formula for the spin-polarization dependence of $1/\tau_{\rm sd}$  in the low temperature regime, which is completely different from what was obtained earlier from an approximate argument~\cite{pustilnik_2003}.
These results provide the necessary input for the application of time-dependent current-spin density functional theory~\cite{Qian03} to the study of the propagation of spin pulses in the nonlinear regime~\cite{gao_preparation}.  This in turn will be useful in future comparisons between theory and experiments on spin-pulse propagation in $1D$ cold Fermi gases.

{\it Note added in proof}---Going beyond second-order perturbation theory, one finds to leading order an additional contribution to $1/\tau_{\rm sd}$ that arises from intra-species interactions. Such a contribution has a power-law temperature dependence~\cite{pustilnik_2003,pustilnik_2006}, and could thus become dominant over the exponential contribution calculated in this work at the lowest temperatures. In Ref.~\onlinecite{pustilnik_2003} it has been shown that this contribution goes as $T^5$, and is proportional to $v^2_0(v_0 - v_{2k_{\rm F}})^2$. This implies that if the interaction range is less than the typical interparticle distance
the fourth-order contribution is expected to be suppressed.

\begin{acknowledgments}
M.P. thanks Michael K\"{o}hl for helpful discussions, and M.P.T. acknowledges the hospitality of the Condensed Matter and Statistical Physics group of the Abdus Salam International Centre for Theoretical Physics in Trieste. We wish to thank 
Michael Pustilnik and Leonid Glazman for useful correspondence. 
G.V. was supported by DOE Grant No. DE-FG02-05ER46203. 
\end{acknowledgments}

\end{document}